\documentclass[aps,pra,twocolumn,showpacs]{revtex4}
\usepackage{color, graphicx}
\usepackage{amsmath, amssymb}
\usepackage{braket}

\definecolor{ROT}{rgb}{0.7,0.1,0.2}

\begin{document}

\title{Witnessing the degree of nonclassicality of light}

\author{M. Mraz}\email{melanie.mraz@uni-rostock.de}\affiliation{Institut f\"ur Physik, Universit\"at Rostock, D-18051 Rostock, Germany}
\author{J. Sperling}\affiliation{Institut f\"ur Physik, Universit\"at Rostock, D-18051 Rostock, Germany}
\author{W. Vogel}\affiliation{Institut f\"ur Physik, Universit\"at Rostock, D-18051 Rostock, Germany}
\author{B. Hage}\affiliation{Institut f\"ur Physik, Universit\"at Rostock, D-18051 Rostock, Germany}

\begin{abstract}
	We introduce an experimentally accessible method to measure a unique degree of nonclassicality, based on the quantum superposition principle, for arbitrary quantum states.
	We formulate witnesses and test a given state for any particular value of this measure.
	The construction of optimal tests is presented as well as the general numerical implementation.
	We apply this approach on examples such as squeezed states, and we show how to formulate conditions to certify a particular degree of nonclassicality for single- and multimode radiation fields.
\end{abstract}
\date{\today}
\maketitle

\section{Introduction}
\label{Sec:1}
	An established way to identify nonclassicality of a quantum state is given by the features of the Glauber-Sudarshan $P$~representation~\cite{glauber63,sudarshan63}.
	It is based on the notion of coherent states to mark the border between quantum and classical physics.  
	A state is nonclassical, if its $P$~function fails to be interpreted as a classical probability~\cite{tit-glau,mandel}.
	Although the $P$~function cannot be measured directly in general, a filtered version of this quasiprobability has been introduced~\cite{KV10} and experimentally observed~\cite{KVBZ11,kieselHage,KVCBAP12}.
	
	Beyond the mere identification, during the last years different attempts were made to quantify nonclassicality.
	One of the early approaches is based on the trace distance of a given quantum state to the set of all classical states~\cite{Hillery1,PolzikEisert,Hillery2}.
	Analogously, a number of distance-based nonclassicality measures were proposed --
	e.g., the Bures distance~\cite{Marian}, or measures based on the Hilbert-Schmidt-norm~\cite{Dodonov1,Dodonov2}.
	An information science-based approach was formulated in terms of the Fisher information~\cite{Hall}.
	Other methods use the occurring negativities of the $P$~function.
	For example, the amount of Gaussian noise which is necessary to remove the negativities of the $P$~function was proposed to quantify the nonclassicality~\cite{Lee1,Lee2,Lutkenhaus}.
	
	Alternatively, a method to quantify the nonclassicality of a quantum state was defined by the potential of the state to generate entanglement~\cite{Asboth}.
	This led from the quantification of nonclassicality to the quantification of entanglement, which is a similarly cumbersome problem.
	One possibility to quantify entanglement is the Schmidt number~\cite{Horodecki09,Guehne,SV11a,SV11}.
	Among other attempts of entanglement quantification, this measure is most closely related to the quantum superposition principle being the foundation of quantum correlations, cf., e.g.,~\cite{RBH01}.
	As the quantification of nonclassicality and of entanglement are similar problems, our idea will adapt the knowledge from entanglement quantification, to quantify the amount of nonclassicality.

	Recently, the amount of nonclassicality for pure and mixed quantum states has been defined from two points of view:
	an operational and an algebraic one, denoted as degree of nonclassicality~\cite{Gerke}.
	The algebraic amount of single mode nonclassicality has been shown to be identical to the amount of entanglement in the output ports of a beam splitter~\cite{VS14}.
	This measure is based on the decomposition of a quantum state in terms of superpositions of coherent states, which resemble the classical harmonic oscillator most closely~\cite{Gazeau}.
	The more superpositions of coherent states are required for the representation of the state under study, the more nonclassical quantum interferences are produced.
	For the notion of entanglement this directly relates to the Schmidt rank for pure states~\cite{NC-book}, or the Schmidt number for mixed ones~\cite{TH00}.

	In the present contribution, we formulate a witness approach in order to determine the degree of nonclassicality.
	This will be done via the formulation and solution of an optimization problem regarding a given number of superpositions of coherent states.
	For the solution of this optimization problem, we will present specific analytical relations and a general numerical approach.
	This allows to formulate accessible constraints to verify a certain degree of nonclassicality.
	We apply our method to different states.
	Moreover, we generalize our approach to witness the degree of nonclassicality in multimode scenarios.

	This paper is structured as follows.
	We introduce the witnessing method in Sec.~\ref{Sec:2} together with the derivation of optimization constraints.
	In Sec.~\ref{Sec:3}, we study analytical and numerical approaches for the witness construction together with experimentally relevant examples of quantum states.
	A generalization to multimode systems is given in Sec.~\ref{Sec:4}.
	Finally, a summary and conclusions are given in Sec.~\ref{Sec:5}.

\section{Witnesses for the degree of nonclassicality}
\label{Sec:2}
	We start with a brief recapitulation of the quantification of nonclassicality.
	Afterwards, we introduce witnesses for the amount of nonclassicality.
	Eventually, we formulate necessary and sufficient conditions for a certain degree of nonclassicality and study the subsequent properties.

\subsection{General definition}
	The main idea for the considered quantification is a decomposition of a quantum state into superpositions of coherent states,
	\begin{align}\label{eq:purestatesdef}
		|\psi_r\rangle=\lambda_1|\alpha_1\rangle+\ldots+\lambda_r|\alpha_r\rangle,
	\end{align}
	where $\lambda_{k}\in\mathbb C\setminus\{0\}$ and $|\alpha_k\rangle$ are coherent states (for $k=1,\ldots,r$).
	The number of superpositions $r$ is our nonclassicality measure for pure states~\cite{Gerke,VS14}.
	Quantum superpositions induce quantum interferences and nonclassical correlations, which can be used as a resource for applications in quantum informations science.
	This is due to the fact that the degree of nonclassicality can be perfectly mapped to the same amount of entanglement in terms of the Schmidt number.

	The set $\mathcal S_r$ denotes the closure of all pure states with a number of superpositions less than, or equal to $r$.
	We aim to give a general nonclassicality measure.
	Hence, we need to consider mixed states as well.
	Therefore, a convex roof construction yields~\cite{Gerke}:
	\begin{align}
		\label{eq:RM}
		\hat{\rho}_{r}=\int_{\ket{\psi_{r}}\in \mathcal S_{r}}{\text{d}P_{\rm cl}(\psi_{r})\ket{\psi_{r}}\bra{\psi_{r}}}.
	\end{align}
	Here, $P_{\rm cl}$ is a classical probability distribution.
	Hence, all these states $\hat{\rho}_{r}$ are elements of the closed, convex set of states $\mathcal M_{r}$,
	\begin{align}
		\mathcal M_{r}=\overline{\text{conv}\left\{
			\ket{\psi_{r}}\bra{\psi_{r}}:\ket{\psi_{r}}\in \mathcal S_{r}
		\right\}},
	\end{align}
	where the closure is performed with respect to the trace norm.
	Now it is possible to properly define a degree of nonclassicality, $D_{\rm Ncl}$, for a quantum state $\hat{\rho}$ by
	\begin{align}\label{eq:inM}
		\hat{\rho}\in \mathcal M_{r} &\Leftrightarrow D_{\rm Ncl}(\hat{\rho})\leq r\\
		\label{eq:notinM}
		\hat{\rho}\notin \mathcal M_{r} &\Leftrightarrow D_{\rm Ncl}(\hat{\rho})>r\\
		\label{eq:inMnotin}
		\hat{\rho}\in \mathcal M_{r}\setminus \mathcal M_{r-1} &\Leftrightarrow D_{\rm Ncl}(\hat{\rho})=r.
	\end{align}
	This means that the degree of nonclassicality is equal to $r$, if and only if $\hat{\rho}$ lies in $\mathcal M_{r}$, but not in $\mathcal M_{r-1}$.

\subsection{Witnessing approach}
	\begin{figure}[ht]
		\centering
		\includegraphics*[width=8cm]{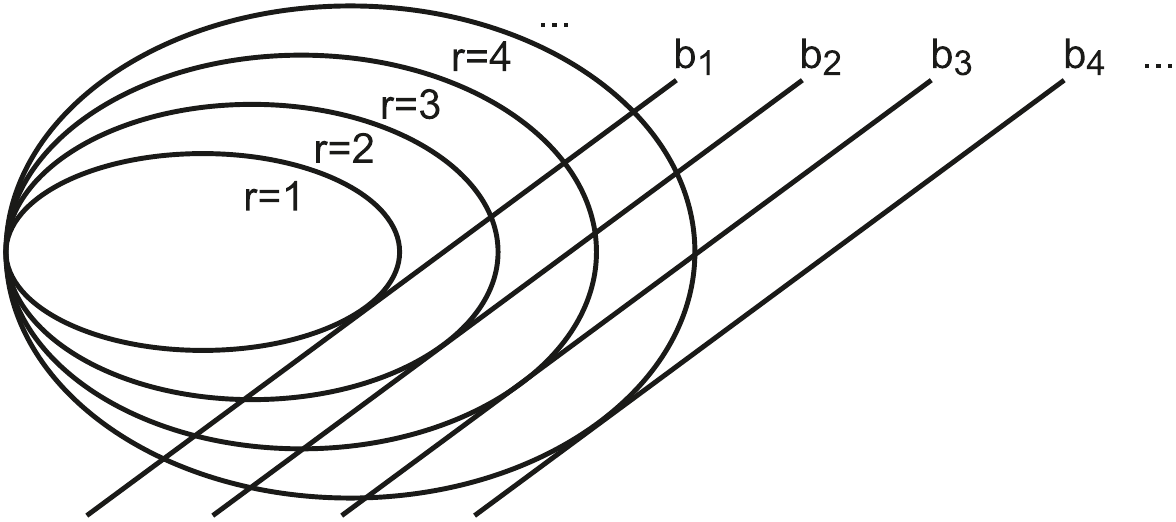}
		\caption{
			Schematic representation of the application of the Hahn-Banach separation theorem.
			The closed, convex, nested sets $\mathcal M_r$ are depicted for several $r$.
			The degree of nonclassicality shall be determined by tangent hyperplanes.
			Here, all states on the right-hand side of this tangent cannot be elements of $\mathcal M_r$.
			By the determination of the parameter $b_r$, it is possible to decide whether a state lies in $\mathcal M_{r}$ or not.
		}\label{fig:HahnBanachMixed}
	\end{figure}

	For witnessing this degree we may apply the Hahn-Banach separation theorem (see, e.g.,~\cite{HBBuch}) as it is visualized in Fig.~\ref{fig:HahnBanachMixed}.
	It allows to separate a closed, convex set and a single state -- not being an element of this set -- from each other.
	The formulation of the theorem in our specific scenario is the following.
	For any state $\hat\varrho\notin \mathcal M_r$ exists a Hermitian operator $\hat{K}$ such that
	\begin{align}\label{eq:construction0}
		\langle \hat K\rangle=&{\rm Tr}(\hat\varrho\hat K)> b_{r}(\hat{K}),\\
		\text{with }b_{r}(\hat{K})=&\sup_{|\psi_r\rangle\in \mathcal S_r} \frac{\langle \psi_r|\hat K|\psi_r\rangle}{\langle \psi_r|\psi_r\rangle}.
		\label{eq:construction1}
	\end{align}
	The least upper bound $b_{r}(\hat{K})$ denotes the maximally attainable expectation value of $\hat K$ for states in $\mathcal M_r$.
	We may formulate this condition in terms of a witness operator
	\begin{align}\label{eq:witnessdefinition}
		\hat W_r=b_{r}(\hat{K})\hat 1-\hat K.
	\end{align}
	This operator has the property that
	\begin{align}
		{\rm Tr}(\hat\rho_r\hat W_r)\geq 0
		\text{ for all }\hat\rho_r\in \mathcal M_r,
	\end{align}
	and ${\rm Tr}(\hat\varrho\hat W_r)<0$, for the considered state $\hat\varrho\notin \mathcal M_r$.
	This implies $D_{\rm Ncl}(\hat{\rho})>r$.

	We obtain, that:
	\begin{align}
	 \rm{Eq}.~\eqref{eq:inM}&~\Leftrightarrow~\forall~\hat W_r:{\rm Tr}(\hat\rho\hat W_r)\geq 0\\
	 \text{ and } \rm{Eq}.~\eqref{eq:notinM}&~\Leftrightarrow~\exists~\hat W_r:\,{\rm Tr}(\hat\rho\hat W_r)< 0.
	\end{align}
	Consequently, we can formulate the following necessary and sufficient condition.
	A quantum state $\hat\varrho$ has a degree of nonclassicality of $r$, $D_{\rm Ncl}(\hat\varrho)=r$, if and only if
	\begin{align*}
		\exists~\hat W_{r-1}:\, {\rm Tr}(\hat\varrho\hat W_{r-1})<0
		\text{ and }\forall~\hat W_{r}:\, {\rm Tr}(\hat\varrho\hat W_{r})\geq 0.
	\end{align*}
	This statement is identical to the definition in Eq.~\eqref{eq:inMnotin}.

	Moreover, let us comment that one could also use the infimum for the construction of a witness,
	\begin{align}\label{eq:witnessMIN}
		\hat W'_r=\hat K-b'_r(\hat K)\hat 1
		\text{ and }b'_r(\hat K)=\inf_{|\psi_r\rangle\in \mathcal S_r} \frac{\langle \psi_r|\hat K|\psi_r\rangle}{\langle \psi_r|\psi_r\rangle}.
	\end{align}
	Thus, we could write for ${\rm Tr}(\hat\varrho\hat W'_r)<0$:
	\begin{align}\label{eq:construction2}
		\langle\hat K\rangle={\rm Tr}(\hat\varrho\hat K)<b'_r(\hat K).
	\end{align}
	This means that the measured expectation value, $\langle\hat K\rangle$, is below the bound $b'_r(\hat K)$, which is the minimal possible expectation value of $\hat K$ for states with a degree of nonclassicality of $r$.
	Due to the nested structure of the sets, we have the general relation:
	\begin{align}
		\nonumber \mathcal M_1\subset\mathcal M_2\subset&\dots\subset\mathcal M_\infty,\\
		b_1(\hat K)\leq b_2(\hat K)&\dots\leq b_\infty(\hat K),\\
		\nonumber b'_1(\hat K)\geq b'_2(\hat K)&\dots\geq b'_\infty(\hat K).
	\end{align}

	There is a well established definition for a degree of entanglement -- the so-called Schmidt number~\cite{Horodecki09}.
	Schmidt-number witness methods have been formulated~\cite{SBL01,SV11a}, being analogous to the nonclassicality approach given above.
	For example, this leads to witnesses which apply to Gaussian states~\cite{SSV13} or in microcavity systems~\cite{PFSV12}.
	In the following, we propose an optimization scheme for the witnesses in Eqs.~\eqref{eq:witnessdefinition} and~\eqref{eq:witnessMIN}.

\subsection{Optimization problem}
	So far, we studied the definition of the degree of nonclassicality and the formal construction of measurable witness operators of this property.
	For the application of the construction scheme of $\hat W_r$ from a general Hermitian operator $\hat K$, see Eqs.~\eqref{eq:construction1} and~\eqref{eq:witnessdefinition}, we need to compute the value of $b_r(\hat K)$.
	This parameter is defined as the least upper bound of the normalized expectation value of $\hat{K}$ for elements in $\mathcal S_r$.
	Hence we have the optimization problem
	\begin{align}\label{eq:SearchOpt}
		b_r=\frac{\bra{\psi_{r}}\hat{K}\ket{\psi_{r}}}{\braket{\psi_{r}|\psi_{r}}}\to \text{optimum},
	\end{align}
	where the optimization is performed with respect to $|\psi_r\rangle\in\mathcal S_r$.
	The maximum $b_r(\hat K)$ of all optima $b_r$ is
	\begin{align}
		b_{r}(\hat K)=\sup\{b_r\}.
	\end{align}
	Similarly, we have for the witness construction in Eq.~\eqref{eq:witnessMIN}: $b'_{r}(\hat K)=\inf\{b_r\}$.
	Using Eq.~\eqref{eq:purestatesdef}, we may rewrite
	\begin{align}\label{eq:bGeneral}
		       b_r=\frac{\sum_{k_1,k_2=1}^{r}{\lambda_{k_1}^\ast\lambda_{k_2}\bra{\alpha_{k_1}}\hat{K}\ket{\alpha_{k_2}}}}{\sum_{k_1,k_2=1}^{r}{\lambda_{k_1}^\ast\lambda_{k_2}\braket{\alpha_{k_1}|\alpha_{k_2}}}}.
	\end{align}

	For finding the least upper bound of this quantity, we can use the necessary optimality conditions
	\begin{align}\label{eq:startDerivation}
		0=\frac{\partial b_r}{\partial \lambda_k^\ast}
		\quad\text{and}\quad
		0=\frac{\partial b_r}{\partial \alpha_k^\ast},
	\end{align}
	for $k=1,\ldots,r$.
	The first equation can be computed as
	\begin{align}\label{eq:optprlimI1}
		0=\frac{\langle \alpha_k|\hat K|\psi_r\rangle}{\langle\psi_r|\psi_r\rangle}-\frac{\langle\psi_r|\hat K|\psi_r\rangle\langle\alpha_k|\psi_r\rangle}{\langle\psi_r|\psi_r\rangle^2}.
	\end{align}
	If we use the definition~\eqref{eq:SearchOpt} of $b_r$, this expression reduces to
	\begin{align}\label{eq:HB1}
		\sum_{l=1}^{r}{\bra{\alpha_{k}}\hat{K}\ket{\alpha_{l}}\lambda_{l}}&=b_r\sum_{l=1}^{r}{\braket{\alpha_{k}|\alpha_{l}}\lambda_{l}}.
	\end{align}
	It is convenient to write Eq.~\eqref{eq:HB1} in a vectorial notion
	\begin{align}\label{eq:HB1vec}
		\boldsymbol G_{\hat K}\boldsymbol\lambda=& b_r\boldsymbol G_{\hat 1}\boldsymbol\lambda,
	\end{align}
	with $\boldsymbol\lambda=(\lambda_l)_{l=1}^r\in\mathbb C^r$ being the optimal coefficients in Eq.~\eqref{eq:purestatesdef} and the matrix
	\begin{align}
		\boldsymbol G_{\hat L}=&(\langle\alpha_k|\hat L|\alpha_l\rangle)_{k,l=1}^r,
	\end{align}
	for operators $\hat L$.
	Useful properties of the map $\hat L\mapsto\boldsymbol G_{\hat L}$ are studied in Appendix~\ref{app:Gmap}.

	We observe that the value $b_r$ corresponds to a generalized eigenvalue of Eq.~\eqref{eq:HB1vec}.
	This also allows us in the following a systematic treatment for solving this problem by applying standard methods for eigenvalue problems.
	In order to find the bounds $\sup\{b_r\}$ and $\inf\{b_r\}$ for an increasing number of possible superpositions $r$, we have to increase the dimensionality of the underlying eigenvalue equation~\eqref{eq:HB1vec}.

	The second conditions in~\eqref{eq:startDerivation} is a little bit more sophisticated.
	First, we recall the relations $\partial_{\alpha^\ast}\langle\alpha|=\langle\alpha|\left(\hat a-\frac{\alpha}{2}\right)$ and $\partial_{\alpha^\ast}|\alpha\rangle=\left(-\frac{\alpha}{2}\right)|\alpha\rangle$.
	Hence, we may rewrite the optimality condition as:
	\begin{widetext}
	\begin{align}
		 0=&\frac{\lambda_k^\ast\langle\alpha_k|\left(\hat a-\frac{\alpha_k}{2}\right)\hat K|\psi_r\rangle+\lambda_k\langle\psi_r|\hat K\left(-\frac{\alpha_k}{2}\right)|\alpha_k\rangle}{\langle \psi_r|\psi_r\rangle}
		-\frac{\langle \psi_r|\hat K|\psi_r\rangle
			\left[\lambda_k^\ast\langle\alpha_k|\left(\hat a-\frac{\alpha_k}{2}\right)|\psi_r\rangle+\lambda_k\langle\psi_r|\left(-\frac{\alpha_k}{2}\right)|\alpha_k\rangle\right]
		}{\langle \psi_r|\psi_r\rangle^2}.
	\end{align}
	\end{widetext}
	Using Eq.~\eqref{eq:optprlimI1}, this expression simplifies to
	\begin{align}\label{eq:HB2}
		0=\lambda_k^\ast\left(\langle\alpha_k|\hat a\hat K|\psi_r\rangle-b_r\langle\alpha_k|\hat a|\psi_r\rangle\right).
	\end{align}
	Without a loss of generality we can assume that $\lambda_k\neq0$, since the case $\lambda_{k_0}=0$ for some $k_0$ would simply correspond to a degree of nonclassicality of $r_0<r$.
	This allows to formulate another vectorial eigenvalue equation
	\begin{align}\label{eq:HB2vec}
		\boldsymbol G_{\hat a\hat K}\boldsymbol\lambda=& b_r\boldsymbol G_{\hat a}\boldsymbol\lambda.
	\end{align}

	For obtaining a physical interpretation of this constrain, we may perform a summation over all $k=1,\ldots,r$ in Eq.~\eqref{eq:HB2}.
	This yields
	\begin{align}\label{eq:dsadasd1}
		\langle \psi_r|\hat a\hat K|\psi_r\rangle=b_r\langle \psi_r|\hat a|\psi_r\rangle.
	\end{align}
	Now we multiply Eq.~\eqref{eq:HB1} with $\alpha_k^\ast$, use $\alpha^\ast\langle \alpha|=\langle \alpha|\hat a^\dagger$.
	A summation over $k$ gives
	\begin{align}\label{eq:dsadasd2}
		\langle \psi_r|\hat a^\dagger\hat K|\psi_r\rangle=b_r\langle \psi_r|\hat a^\dagger|\psi_r\rangle.
	\end{align}
	Finally, the difference of Eq.~\eqref{eq:dsadasd1} and the conjugated Eq.~\eqref{eq:dsadasd2} reads as
	\begin{align}\label{eq:NoCommutator}
		\langle\psi_r|[\hat a,\hat K]|\psi_r\rangle=0.
	\end{align}
	Note that this condition may be alternatively written in vectorial notion as $0=\boldsymbol \lambda^\dagger\boldsymbol G_{[\hat a,\hat K]}\boldsymbol \lambda$.
	Although the derivation of this condition was not so trivial, its physical interpretation is quite surprising.
	The optimal state $|\psi_r\rangle$ has a vanishing mean value of the quantum mechanical commutator of the observable $\hat K$ and the field component $\hat a$.

\subsection{Transformation properties}
	Useful characteristics of our equations are transformation properties.
	For example one could use an operator $\hat K'=\mu\hat 1+\nu\hat K$ instead of $\hat K$.
	Using the properties of the map $\boldsymbol G$, it turns out that the eigenvalues exhibit the same transformed structure,
	\begin{align}\label{eq:rescaleNshift}
		\boldsymbol G_{\hat K'}=\mu\boldsymbol G_{\hat 1}+\nu\boldsymbol G_{\hat K}
		\quad\Rightarrow\quad b_r'=\mu+\nu b_r.
	\end{align}
	We could also consider an operator
	\begin{align}\label{eq:displacement}
		\hat K_\beta=\hat D(\beta)\hat K\hat D(\beta)^\dagger,
	\end{align}
	where $\hat D(\beta)=\exp[\beta\hat a^\dagger-\beta^\ast\hat a]$ is the displacement operator.
	This displaced operator has the same extremal values as the initial operator $b_{r,\beta}=b_r$, whereas the optimal state $|\psi_r\rangle$ is decomposed in terms of displaced coherent states $\hat D(\beta)|\alpha_k\rangle$ ($k=1,\ldots,r$).
	Analogously to the displacement, we can perform a phase rotation:
	\begin{align}\label{eq:phaserotation}
		\hat K_\varphi=\exp[-{\rm i}\varphi\hat n]\hat K\exp[{\rm i}\varphi\hat n],
	\end{align}
	again the extremal values remain unperturbed, $b_{r,\varphi}=b_r$, and the coherent states are rotated in phase space,
	$\exp[-{\rm i}\varphi\hat n]|\alpha_j\rangle=|\exp[-{\rm i}\varphi]\alpha_j\rangle$.

	More generally, one can show with some simple algebra that any operator $\Lambda$ with the property $\Lambda(|\alpha\rangle\langle\alpha|)=|\alpha'\rangle\langle\alpha'|$, for an invertible function $\alpha'=f(\alpha)$, does not change the optimal values $b_r$.
	One example is the transposition, $(|\alpha\rangle\langle \alpha|)^{\rm T}=|\alpha^\ast\rangle\langle \alpha^\ast|$.
	Such transformation properties are useful for the construction of a whole class of witnesses from a single one.
	The considered operations do not change the optimal values $b_r$, they transform them in a unique form, for example, Eq.~\eqref{eq:rescaleNshift}.
	Therefore, the optimization problem needs to be solved for only one element of a complete class of Hermitian operators.

\subsection{Preliminary results}
	We may summarize our preliminary findings.
	A method to witness the degree of nonclassicality has been formulated in terms of Hermitian operators $\hat K$.
	The corresponding bounds $b_r(\hat K)$ and $b'_r(\hat K)$ for $r\in\mathbb N$ are given by the maximal or minimal eigenvalue of Eq.~\eqref{eq:HB1vec}, respectively.
	These bounds give the maximal or minimal expectation value of $\hat K$ for all elements in the set $\mathcal M_r$.
	Whenever the expectation value, $\langle\hat K\rangle={\rm Tr}(\hat\rho\hat K)$, exceeds the upper or lower bound, we find that $\hat\rho\notin\mathcal M_r$, i.e., $D_{\rm Ncl}(\hat\rho)>r$.
	The optimal state $|\psi_r\rangle=\sum_{k=1}^r\lambda_r|\alpha_r\rangle$ fulfills the optimization constrains formulated by the eigenvalue problem in Eq.~\eqref{eq:HB1vec}, and it exhibits a vanishing mean value of the commutator in Eq.~\eqref{eq:NoCommutator}.
	Additionally, we studied useful transformation properties of our approach.

\section{Solutions, Numerical Implementation, and Examples}
\label{Sec:3}
	Let us now apply our method to some examples.
	We will consider analytical and numerical solutions for different degrees of nonclassicality $r$.
	Some examples are devoted to find necessary and sufficient witnesses for pure states.

\subsection{Formal Solutions}
	In the case $r=1$, the optimization condition in Eq.~\eqref{eq:HB1vec} simplifies to $b_1=\langle \alpha_1|\hat K|\alpha_1\rangle$.
	We may also study an operator $\hat f=f(\hat a^\dagger,\hat a)$, which is a function $f$ of annihilation $\hat a$ and creation operators $\hat a^\dagger$.
	We define
	\begin{align}\label{eq:NormallyOrdered}
		\hat K={:}\hat f^\dagger\hat f{:}
		\text{ and }
		b_r=|f(\alpha_1^\ast,\alpha_1)|^2\geq0,
	\end{align}
	where ${:}\,\cdot\,{:}$ denotes the normal ordering prescription.
	In this case we get from our approach -- using the witness construction in Eq.~\eqref{eq:witnessMIN} including the minimal eigenvalue $b_1$ -- the consistent construction of witnesses for nonclassicality:
	\begin{align}
		\hat W_1=\hat K-0={:}\hat f^\dagger\hat f{:},
	\end{align}
	cf., e.g., Ref.~\cite{SV05}.
	Note that in the case, $b'_r(\hat K)=\inf_{\alpha_1}|f(\alpha_1^\ast,\alpha_1)|^2>0$, the witness $\hat W_1$ is not optimal.
	This means that the witness $\hat W_1^{\rm (opt)}={:}\hat f^\dagger\hat f{:}-b'_r({:}\hat f^\dagger\hat f{:})\hat 1$ is even finer than $\hat W_1$, see Ref.~\cite{LKCH00} for the equivalent definition of finer or optimal entanglement witnesses.

	Let us continue with the case $r>1$.
	For $r=2$, the vectorial form in Eq.~\eqref{eq:HB1vec} is an eigenvalue problem of a $2\times2$ matrix.
	The solutions for such a problem are known, and here they read as
	\begin{align}
		b_{2}^{\pm}=&\frac{1}{2}\left[{\rm Tr}(\boldsymbol G_{\hat 1}^{-1}\boldsymbol G_{\hat K})\pm\Delta\right],\\
		\Delta=&\sqrt{\left[{\rm Tr}(\boldsymbol G_{\hat 1}^{-1}\boldsymbol G_{\hat K})\right]^2-4\det\left[\boldsymbol G_{\hat 1}^{-1}\boldsymbol G_{\hat K}\right]}.
	\end{align}
	More generally, $r>2$, the eigenvalue problem in~\eqref{eq:HB1vec} has no such simple solution.
	In such a scenario, one has to find the roots of the characteristic polynomial.
	The characteristic polynomial of the eigenvalue problem~\eqref{eq:HB1vec} reads as
	\begin{align}\label{eq:RootChracteristicPoly}
		0=&\chi(b_r)=\det\left[\boldsymbol G_{\hat K}-b_r\boldsymbol G_{\hat 1}\right].
	\end{align}
	In this general case, the minimal or maximal root $b_r$ of this polynomial for arbitrary choices $\alpha_1,\ldots,\alpha_r$ yields the value of $b'_r(\hat K)$ or $b_r(\hat K)$, respectively.

\subsection{General numerical implementation}
\label{sec:GenImple}

	Based on Eq.~\eqref{eq:RootChracteristicPoly}, we can formulate a proper numerical implementation.
	This is given by the following approach:
	\begin{itemize}
		\item[(i)] compute the minimal/maximal root of $\chi$;
		\item[(ii)] minimize/maximize this root over the choice of $(\alpha_1,\ldots,\alpha_r)\in\mathbb C^r$.
	\end{itemize}
	This general method allows to construct the bounds for any measured observable $\hat K$.
	Then the method has to be applied as follows.
	The experiment yields the expectation value $\langle \hat K\rangle$.
	This value can be compared with the bound for arbitrary degrees of nonclassicality $r$, see Eqs.~\eqref{eq:construction0} and~\eqref{eq:construction2}.

	As an example, let us consider a witness based on quadrature variances: $\hat K=\left[\Delta\hat x(\varphi)\right]^2$.
	Note that this operator is only bounded from below.
	Due to the displacement invariance, cf. Eq.~\eqref{eq:displacement}, and the phase rotation invariance, cf. Eq.~\eqref{eq:phaserotation}, we can -- without loss of generality -- restrict our considerations to
	\begin{align}
	\label{eq:quadop}
		\hat K=\hat x(0)^2=(\hat a+\hat a^\dagger)^2=2\hat a^\dagger\hat a+\hat a{}^2+\hat a^\dagger{}^2+\hat 1.
	\end{align}
	First, for $r=1$ the quadrature variance of coherent states is bounded from below by one, $b'_1(\hat K)=1$.
	Second, for arbitrary states ($r=\infty$) we have a minimum $b'_\infty(\hat K)=0$, obtained by infinitely squeezed states.

	We summarize the numerically obtained boundaries for some values $r$ in Table~\ref{tab:Numerics}.
	Additionally, the corresponding bounds to the squeezing power are given.
	For states with a quadrature variance below the boundary $b'_r(\hat K)$, the degree of nonclassicality is larger than $r$.
	The chosen observable yields a clear relation between the observed degree of nonclassicality and the squeezing strength.
	For the so far strongest realized squeezing of 12.7~dB~\cite{WRS}, the corresponding degree of nonclassicality is $r=8$, cf. Table~\ref{tab:Numerics}.

\begin{table}[ht!]
	\caption{
		Minimal expectation values $b_r'(\hat K)$ for states in the set $\mathcal M_r$ of the observable $\hat K=\left[\Delta\hat x(\varphi)\right]^2$ are listed.
		Whenever the squeezing power of the experimentally realized state $\hat\rho$ exceeds the squeezing bounds, we have $D_{\rm Ncl}(\hat \rho)>r$.
	}\label{tab:Numerics}
	\begin{center}
	\begin{tabular}{ c p{1cm} c p{1cm} c}
		\hline\hline
		$r$ & & $b_r'(\hat K)$ & & squeezing\\
		\hline
		$1$ & & $1.000000$ & & $0.00$~dB \\
		$2$ & & $0.443071$ & & $3.54$~dB \\
		$3$ & & $0.256447$ & & $5.91$~dB \\
		$4$ & & $0.169295$ & & $7.71$~dB \\
		$5$ & & $0.121006$ & & $9.17$~dB \\
		$6$ & & $0.091245$ & & $10.4$~dB \\
		$7$ & & $0.071510$ & & $11.4$~dB \\
		$8$ & & $0.057702$ & & $12.4$~dB \\
		$9$ & & $0.047638$ & & $13.2$~dB \\
		$\vdots$ & & $\vdots$ & & $\vdots$\\
		$\infty$ & & $0$ & & $\infty$\\
		\hline\hline
	\end{tabular}
	\end{center}
\end{table}

\subsection{Pure states}
\label{Sec:PureStates}
	In order to check whether a pure state is of a particular degree of nonclassicality, one may compute its distance $d_r$ to the set $\mathcal S_r$,
	\begin{align}
		\nonumber d_r=&\left\|\ket{\psi}-\frac{|\psi_r\rangle}{\sqrt{\langle\psi_r|\psi_r\rangle}}\right\|^2
		\\=&{\rm Tr}\left(\ket{\psi}\bra{\psi}-\frac{|\psi_r\rangle\langle\psi_r|}{\langle\psi_r|\psi_r\rangle}\right)^{2}\to\min,
	\end{align}
	where $\ket{\psi}$ is the quantum state under study and $|\psi_r\rangle$ is a coherent superposition state with a known degree of nonclassicality.
	This expression -- taking the normalizations into account -- is
	\begin{align}\label{eq:distancewitness}
		d_r=2\left[1-\frac{\langle\psi_r|\left(|\psi\rangle\langle\psi|\right)|\psi_r\rangle}{\langle\psi_r|\psi_r\rangle}\right].
	\end{align}
	A careful look on this distance yields the proper choice for the construction of a witness, namely:
	\begin{align}
		\hat K=|\psi\rangle\langle\psi|.
	\end{align}
	The value of $b_{r}(\hat K)$ gives the information which of the cases \eqref{eq:inM}--\eqref{eq:inMnotin} holds true.
	If $b_{r}(\hat K)$ is equal to one, $d_r=0$, then the state $|\psi\rangle\langle\psi|$ lies in the set $\mathcal S_{r}$.
	If $b_{r}(\hat{K})$ is smaller than one, $d_r>0$, then the state is not in the set $\mathcal S_{r}$.
	Combining both facts, we observe that we found an optimal, necessary, and sufficient witness for arbitrary pure state $|\psi\rangle$:
	\begin{align}\label{eq:Wit}
		\hat W_r=b_r(|\psi\rangle\langle\psi|)\hat 1-|\psi\rangle\langle\psi|.
	\end{align}
	It is worth mentioning that $\langle \hat W_r\rangle<0$ also detects a degree of nonclassicality beyond pure state.
	However, in the mixed state case $\langle \hat W_r\rangle\geq0$ does not imply a degree less or equal to $r$.
	Since for mixed state this witness might not be the best choice.

	Now, let us compute the value $b_r(\hat K)$ for $\hat K=|\psi\rangle\langle\psi|$.
	This means that we have to solve
	\begin{align}
		\boldsymbol G_{|\psi\rangle\langle\psi|}\boldsymbol \lambda=b_r\boldsymbol G_{\hat 1}\boldsymbol \lambda.
	\end{align}
	Since we have a rank one operator, $\boldsymbol G_{|\psi\rangle\langle\psi|}=\boldsymbol g_{|\psi\rangle}\boldsymbol g_{|\psi\rangle}^\dagger$ with $\boldsymbol g_{|\psi\rangle}=(\langle\alpha_i|\psi\rangle)_{i=1}^r$, we get the maximal solution for
	\begin{align}\label{eq:PureSolution}
		\boldsymbol \lambda=\boldsymbol G_{\hat 1}^{-1}\boldsymbol g_{|\psi\rangle}
		\text{ and }
		b_r=\boldsymbol g_{|\psi\rangle}^\dagger \boldsymbol G_{\hat 1}^{-1}\boldsymbol g_{|\psi\rangle}.
	\end{align}

\subsubsection{Finite superposition states}
	For constructing the witness in Eq.~\eqref{eq:Wit}, let us consider finite superposition states, $R\geq r$,
	\begin{align}\label{eq:finitesuper}
		|\psi\rangle=\sum_{k=1}^R \kappa_k |\beta_k\rangle.
	\end{align}
	Let us comment, that one can show a general relation for the case $|\beta_{k_1}-\beta_{k_2}|\gg 1$ (for all $k_1\neq k_2$) in~\eqref{eq:finitesuper}.
	Then, we get almost orthogonal vectors $\langle\beta_{k_1}|\beta_{k_2}\rangle\approx0$.
	This leads to a maximal solution~\eqref{eq:PureSolution} for the finite superposition state,
	\begin{align}\label{eq:Br}
		b_r(|\psi\rangle\langle\psi|)\approx\max\{|\kappa_{k_1}|^2+\ldots+|\kappa_{k_r}|^2\},
	\end{align}
	where the maximum is taken over all pairwise different indices, $k_i\neq k_{i'}$.
	See also the related method for Schmidt number witnesses~\cite{SV11a}.
	
\begin{figure}[ht]
	\includegraphics*[width=7cm]{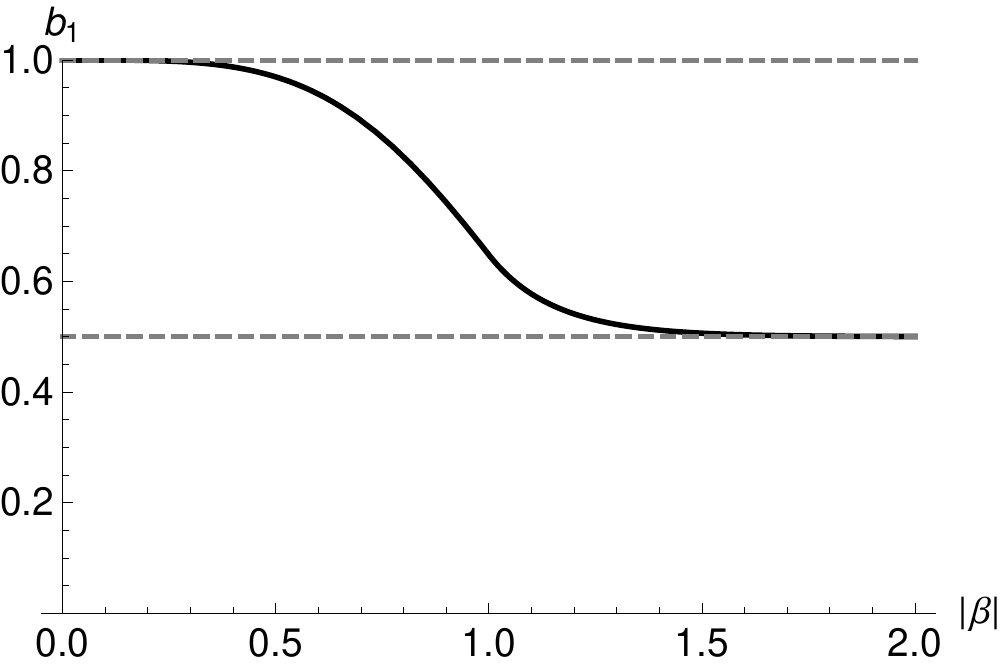}
		\caption{
			The maximal projection, $b_1=b_1(|\psi\rangle\langle\psi|)$ (solid line), of the even coherent state with the set of states with a minimal degree of nonclassicality $\mathcal S_1$ is shown in dependence of the coherent amplitude $|\beta|$ of the coherent components of $|\psi\rangle$.
			The dashed lines depict the limiting values $1$ and $0.5$ for $|\beta|\to 0$ and $|\beta|\to\infty$, respectively.
		}\label{fig:ECS}
	\end{figure}

	As an example, we may study the compass state~\cite{CompState},
	\begin{align}\label{eq:compass}
	  \beta_k=&\beta e^{\frac{2\pi i}{R}k},\\
	  \nonumber \kappa_k=&\left(\sum_{k_1,k_2=1}^{R} \exp\left[-|\beta|^2+|\beta|^2~e^{\frac{2\pi i}{R}(k_2-k_1)}\right] \right)^{-1/2}.
	\end{align}
	being a generalization of the even coherent state for $R=2$~\cite{EOCS,ECS2}.
	In Fig.~\ref{fig:ECS}, we plot $b_1=b_{1}(|\psi\rangle\langle\psi|)$ for the case $R=2$ depending on the $|\beta|$, which is the separation between the two components, $|\beta\rangle$ and $|-\beta\rangle$, of the even coherent state.
	We observe that the overlap with the set of classical states $\mathcal S_1$ is quite large, $b_1\approx 1$, for small coherent amplitudes, and it saturates for $|\beta|\to\infty$ at the expected value $b_1=0.5$.
	
	In general, the compass state in~\eqref{eq:compass} contains $R$ coherent superpositions of coherent states where each state has a certain phase rotation.
	Hence for the compass state holds for $b_{r}$ in Eq.~\eqref{eq:Br} with the coefficients in Eq~\eqref{eq:compass}, in the limit of infinite amplitudes $|\beta|\to\infty$:
	\begin{align}
		b_r=b_r\left(\lim_{|\beta|\to\infty} |\psi\rangle\langle\psi|\right)=\frac{r}{R}
		\text{ for }r<R,
	\end{align}
	and the value $b_r=1$ for $r\geq R$.
	This defines the threshold for the amount of nonclassicality $r$, witnessed by the compass state.
	This results in the bound $b_1=0.5$ in Fig.~\ref{fig:ECS} for $R=2$.

\subsubsection{Infinite superposition states}
	Second let us address the more general case in the Fock basis expansion
	\begin{align}
		|\psi\rangle=\sum_{n=0}^\infty \frac{\psi_n}{\sqrt{n!}} \hat a^\dagger{}^n|{\rm vac}\rangle,
	\end{align}
	for the rank one test operator $\hat K=\left|\psi\rangle\langle\psi\right|$.
	Since the squeezed vacuum states have a large number of applications, it is interesting to investigate in particular their strength of nonclassicality,
	\begin{align}
		\text{for even $n$: } \psi_n=&\frac{1}{\sqrt\mu}\left(-\frac{\nu}{2\mu}\right)^{n/2}\frac{\sqrt{n!}}{(n/2)!}, \\
		\text{for odd $n$: } \psi_n=&0,
	\end{align}
	where $\mu=\cosh(\xi)$ and $\nu=e^{i\arg{\xi}}\,\sinh(\xi)$, or $|\psi\rangle=(1/\sqrt{\mu})e^{-\nu\hat{a}^\dagger{}^2/2\mu}\ket{\rm vac}$.
	From a minimization of the solution in Eq.~\eqref{eq:PureSolution}, we can now compute the projection of the squeezed vacuum state onto arbitrary subsets $\mathcal S_r$.
	Therefore it is useful to take a look on the inner product of the coherent state and the squeezed vacuum state~\cite{VogelWelsch}, as it is used in the calculation, $\braket{\alpha|\xi,0}=e^{-|\alpha|^{2}/2-\nu\alpha^{\ast 2}/(2\mu)}/\sqrt\mu$.

	\begin{figure}[ht]
	\includegraphics*[width=7cm]{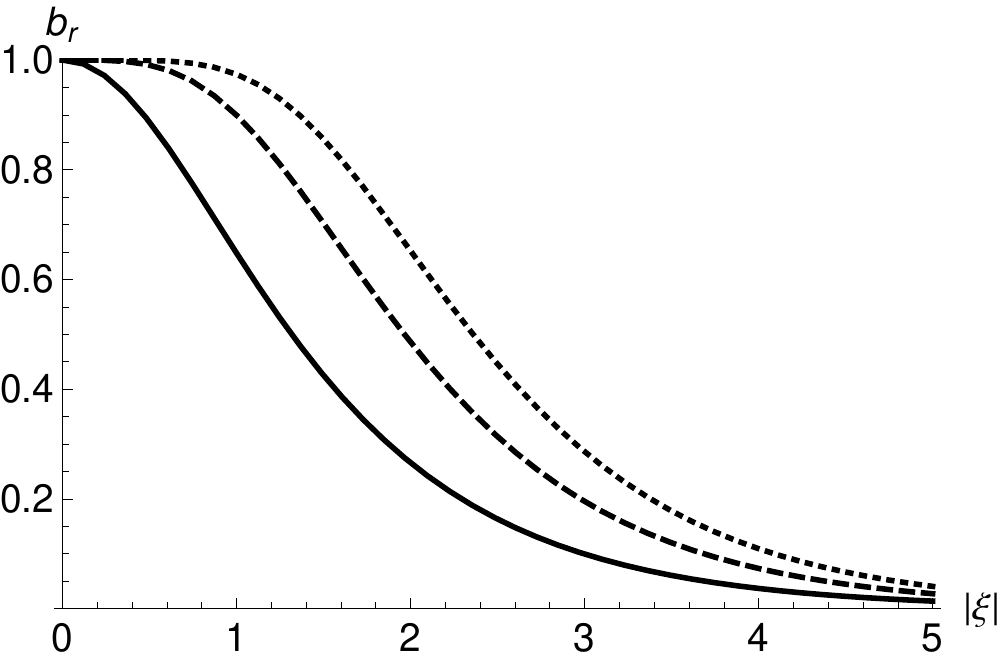}
		\caption{
			The bounds $b_r=b_{r}(|\psi\rangle\langle \psi|)$ ($r=1$ solid, $r=2$ dashed, $r=3$ dotted) are shown in dependence of the squeezing parameter $\xi$, which defines the squeezed vacuum state $|\psi\rangle$.
			We find that a stronger squeezing yields a smaller bound to be violated to verify a degree of nonclassicality above $r$.
		}\label{fig:sqz}
	\end{figure}

	In Fig.~\ref{fig:sqz}, we show the computed bounds $b_r=b_{r}(|\psi\rangle\langle \psi|)$.
	Let us stress that this bound means that whenever the fidelity $\langle\psi|\hat\rho|\psi\rangle$ exceeds the value of $b_r$, we have successfully proven that $D_{\rm Ncl}(\hat\rho)>r$.
	It is worth mentioning, that the results in Table~\ref{tab:Numerics} are directly given for the quadrature operator in Eq.~\eqref{eq:quadop}.
	For the here studied fidelity, it can be seen that a high squeezing clearly yields a low overlap with states in $\mathcal S_r$.
	Additionally, we observe that no finite $r$ and $|\xi|>0$ gives the value $b_{r}=1$, see also~\cite{Gerke}.
	This means that the strength of the nonclassicality of a squeezed vacuum state is independent of the amount of squeezing infinite, $D_{\rm Ncl}(|\psi\rangle\langle\psi|)=\infty$.
	However, it becomes harder to exceed the bounds $b_r$ for $|\xi|$ approaching zero.
	From the experimental point of view, it is intuitive that an increasing squeezing in our system results in an increasing verifiable degree of nonclassicality.
	The expected result is that, due to finite measuring time and statistical fluctuations, we can certify a certain number of quantum superposition.

\section{Multimode Nonclassicality}
\label{Sec:4}
	Similarly to the single mode approach, the $N$-mode degree of nonclassicality may be witnessed.
	For this purpose, we define for pure states a degree of nonclassicality, $D_{\rm Ncl}(|\psi_{r,N}\rangle\langle\psi_{r, N}|)=r$, by the number of coherent superpositions of multimode coherent states,
	\begin{align}\label{eq:MultimodeDec}
		|\psi_{r,N}\rangle=\sum_{j=1}^r \lambda_j |\boldsymbol \alpha_j\rangle,
	\end{align}
	with $\lambda_j\in\mathbb C\setminus\{0\}$ and coherent amplitudes $\boldsymbol \alpha_j\in\mathbb C^N$ ($\boldsymbol \alpha_j\neq \boldsymbol \alpha_{j'}$ for $j\neq j'$).
	A convex roof construction yields the proper multimode nonclassicality measure for mixed states.
	A major advantage of this notion for the degree of nonclassicality is its invariance under classical mode-transformations,
	\begin{align}
		\hat a'_n=\sum_{n'=1}^N U_{n,n'}\hat a_{n'},
	\end{align}
	for $n=1,\ldots,N$ and a unitary matrix $\boldsymbol U=(U_{n,n'})_{n,n'=1}^N$.
	This transformation maps coherent amplitudes as $\boldsymbol\alpha'_j=\boldsymbol U\boldsymbol \alpha_j$, and, therefore, the structure of Eq.~\eqref{eq:MultimodeDec} remains invariant.
	In particular, we have $r'=r$.

	Let us briefly outline how to construct the corresponding witnesses for $D_{\rm Ncl}$ from multimode, Hermitian operators $\hat K$:
	\begin{align}
		\hat W_{r,N}=&b_{r,N}(\hat K)\hat 1-\hat K,\\
		\text{with }b_{r,N}(\hat K)=&\sup_{|\psi_{r,N}\rangle}\frac{\langle \psi_{r,N}|\hat K|\psi_{r,N}\rangle}{\langle \psi_{r,N}|\psi_{r,N}\rangle},
	\end{align}
	or, equivalently,
	\begin{align}
		\hat W'_{r,N}=&\hat K-b'_{r,N}(\hat K)\hat 1,\\
		\text{with }b'_{r,N}(\hat K)=&\inf_{|\psi_{r,N}\rangle}\frac{\langle \psi_{r,N}|\hat K|\psi_{r,N}\rangle}{\langle \psi_{r,N}|\psi_{r,N}\rangle}.
	\end{align}
	The values of $b_{r,N}(\hat K)$ or $b'_{r,N}(\hat K)$ are given by the least upper or smallest lower bound of eigenvalues $b_{r,N}$ of the equation in $\mathbb C^r$:
	\begin{align}
		\boldsymbol G_{\hat K}\boldsymbol \lambda=b_{r,N}\boldsymbol G_{\hat 1}\boldsymbol \lambda,
	\end{align}
	with $\boldsymbol G_{\hat L}=(\langle \boldsymbol \alpha_{i}|\hat L|\boldsymbol \alpha_{j}\rangle)_{i,j=1}^r$.

	The solutions and numerical implementation can be done similarly to the single mode case.
	Let us also note that the multimode Schmidt number~\cite{EB01} for an entangled state $\hat \rho$ and a fixed mode decomposition is smaller or equal then $D_{\rm Ncl}(\hat\rho)=r$, cf.~\cite{VS14}.
	Hence, the amount of entanglement is bounded from above by the multimode degree of nonclassicality $r$ which can be obtained from our witnessing approach.

\section{Summary and Conclusions}
\label{Sec:5}
	In conclusion, we introduced witnesses to measure the amount of nonclassicality in quantum systems.
	This measure is based on the decomposition of any state into coherent superpositions of coherent states.
	For proving that the witnessing approach is necessary and sufficient, we applied the Hahn-Banach separation theorem.
	With this knowledge it has been possible to formulate optimization equations for the estimation of the amount of nonclassicality.
	These equations represent an eigenvalue problem.
	Furthermore different transformation properties were investigated in order to solve the equations only once and to get the corresponding bounds for a whole class of operators.
	After studying the properties of the measure and constructing general optimal witness, the problem of finding the proper witnesses was completely solved for pure states.
	Based on the eigenvalue equation structure, a general numerical algorithm for constructing witnesses was proposed and implemented.
	As an example, we studied an unbounded operator to measure the degree of nonclassicality in terms of squeezing.
	Afterwards we used our method to determine the amount of nonclassicality of different examples for pure states with different complexities.
	The presented approach has been generalized to determine the amount of nonclassicality in multimode radiation fields.
	The relation to witnesses for the amount of entanglement, in terms of the Schmidt number, was also considered.

	The general nonclassicality measure is given by the number of superimposed coherent states.
	The surprising feature of quantum superpositions have been demonstrated in various experiments in quantum optics.
	The presented measure is not only theoretically accessible, i.e. a computable measure.
	By applying our results to experiments, it even becomes a measurable measure of quantumness.
	In multimode fields, the difficulty lies in the fact that field components can be superimposed in addition to quantum superpositions of states.
	We consistently took this fact into account.
	This was done in a way, that our criteria are solely sensitive to quantum interferences.
	Hence, the available amount of quantumness in different optical system can be determined, e.g., for possible applications in quantum technologies.

\section*{Acknowledgments}
	This work was supported by the Deutsche Forschungsgemeinschaft through SFB 652 (B12 and B13).

\appendix
\section{Ring homomorphism $\boldsymbol G$}\label{app:Gmap}
	In the following, we will provide some properties of the map $\hat L\mapsto \boldsymbol G_{\hat L}$.
	This map is defined for a set of coherent states $\{|\alpha_1\rangle,\ldots,|\alpha_r\rangle\}$ as
	\begin{align}
		\boldsymbol G_{\hat L}=\left(\langle\alpha_i|\hat L|\alpha_j\rangle\right)_{i,j=1}^r.
	\end{align}
	This means for the matrix components $(\boldsymbol G_{\hat L})_{i,j}=\langle\alpha_i|\hat L|\alpha_j\rangle$, which can be used to prove all following properties.
	In order to prove the ring homomorphism property, let us define a proper product of two operators:
	\begin{align}
		\hat L_1\circ\hat L_2=\hat L_1\hat Q\hat L_2,
		\text{ with } \hat Q=\sum_{k=1}^r|\alpha_k\rangle\langle\alpha_k|.
	\end{align}
	The standard operator product will be written in the usual form, i.e., without an extra symbol, $\hat L_1\hat L_2$.
	Obviously, $\circ$ is associative.
	Hence, we get the homomorphism properties from
	\begin{align}
		\boldsymbol G_{\mu_1\hat L_1+\mu_2\hat L_2}=&\mu_1\boldsymbol G_{\hat L_1}+\mu_2\boldsymbol G_{\hat L_2},\\
		\boldsymbol G_{\hat L_1\circ\hat L_2}=&\boldsymbol G_{\hat L_1}\boldsymbol G_{\hat L_2}.
	\end{align}
	Moreover, the conjugation property is conserved by this continuous map:
	\begin{align}
		\boldsymbol G_{\hat L^\dagger}=&\left(\boldsymbol G_{\hat L}\right)^\dagger,\\
		\boldsymbol G_{\lim_{n\to\infty}\hat L_n}=&\lim_{n\to\infty}\boldsymbol G_{\hat L_n},
	\end{align}
	which makes $\boldsymbol G$ even a $C{}^\ast$-algebra homomorphism,
	\begin{align}
		\boldsymbol G: {\rm Lin}(\mathcal H\to\mathcal H)\to {\rm Lin}(\mathbb C^r\to\mathbb C^r),
	\end{align}
	with the set ${\rm Lin}(X\to Y)$ denoting the corresponding linear and (typically) bounded operators.

	In order to use the map $\boldsymbol G$ efficiently, let us consider additional properties of this calculus:
	\begin{align}
		\boldsymbol G_{\hat 1}\boldsymbol G_{\hat L}=&\boldsymbol G_{\hat Q\hat L};\\
		\boldsymbol G_{\hat L}\boldsymbol \lambda=&\left(\langle \alpha_i|\hat L|\psi_r\rangle\right)_{i=1}^r,
	\end{align}
	with $\boldsymbol \lambda=\left(\lambda_j\right)_{j=1}^r\in\mathbb C^{r}$ and $|\psi_r\rangle=\sum_{j=1}^r\lambda_j|\alpha_j\rangle\in{\rm span}\{|\alpha_1\rangle,\ldots,|\alpha_r\rangle\}\subset\mathcal H$;
	\begin{align}
		\boldsymbol G_{\hat L\hat a}=\boldsymbol G_{\hat L}\boldsymbol A
		\text{ and }
		\boldsymbol G_{\hat a^\dagger \hat L}=\boldsymbol A^\ast\boldsymbol G_{\hat L},
	\end{align}
	with $\boldsymbol A={\rm diag}(\alpha_1,\ldots,\alpha_r)$.
	The Gram-Schmidt matrix of the studied set of coherent states is $\boldsymbol G_{\hat 1}$.
	The pseudo-inverse $\hat Q^+$ -- i.e., $\hat Q^+\hat Q=\hat P_{\alpha_1,\ldots,\alpha_r}$ being the projector to the subspace ${\rm span}\{|\alpha_1\rangle,\ldots,|\alpha_r\rangle\}$ and $\boldsymbol G_{\hat 1}=\boldsymbol G_{\hat P_{\alpha_1,\ldots,\alpha_r}}$ -- has the property of a unity:
	\begin{align}
		\boldsymbol G_{\hat L}\boldsymbol G_{\hat Q^{+}}=\boldsymbol G_{\hat L\hat Q\hat Q^{+}}=\boldsymbol G_{\hat L\hat P_{\alpha_1,\ldots,\alpha_r}}=\boldsymbol G_{\hat L}.
	\end{align}
	Moreover, we find for rank one operators the decomposition:
	\begin{align}
		\boldsymbol G_{|\psi_2\rangle\langle\psi_1|}=\boldsymbol g_{|\psi_2\rangle}\boldsymbol g_{|\psi_1\rangle}^\dagger,
	\end{align}
	with the definition $\boldsymbol g_{|\psi_{1(2)}\rangle}=(\langle\alpha_i|\psi_{1(2)}\rangle)_{i=1}^r$.
	We also get the result
	\begin{align}
		\boldsymbol G_{\hat L}\boldsymbol g_{|\psi\rangle}=\boldsymbol g_{\hat L \hat Q|\psi\rangle}=\boldsymbol g_{\hat L \circ|\psi\rangle}.
	\end{align}
	It is important to mention that all the listed properties a are also valid in the multipartite case, $\boldsymbol G_{\hat L}=(\langle \boldsymbol \alpha_i|\hat L|\boldsymbol \alpha_j\rangle)_{i,j=1}^r$, with $\boldsymbol \alpha_j\in{\mathbb C}^N$ for $j=1\ldots N$.


\end{document}